\documentstyle[psfig]{mn}

\def\refitem{\par\parskip 0pt\noindent\hangindent 20pt}
\def\lesssim{\mathrel{\hbox{\rlap{\hbox{\lower4pt\hbox{$\sim$}}}\hbox{$<$}}}}
\def\gtrsim{\mathrel{\hbox{\rlap{\hbox{\lower4pt\hbox{$\sim$}}}\hbox{$>$}}}}

\title[Spectral energy distributions of FR~I nuclei and
the FR~I/BL Lac unifying model]       
{Spectral energy distributions of FR~I nuclei  and
the FR~I/BL Lac unifying model}

\author[Alessandro Capetti et al.]
{A. Capetti$^1$, E. Trussoni$^1$, A. Celotti$^2$,
L. Feretti$^3$, M. Chiaberge$^2$\\
$^1$Osservatorio Astronomico di Torino, Strada Osservatorio 20,
I-10025 Pino Torinese, Italy\\
$^2$S.I.S.S.A., Via Beirut 2--4, I-34014 Trieste, Italy\\
$^3$Istituto di Radioastronomia del C.N.R., Via  Gobetti 101, I-40129 
Bologna, Italy}
\date{Received ***; in original form ***}       
\begin{document} 
\maketitle 

\begin{abstract}

We consider archival ROSAT and Hubble Space Telescope (HST)
observations of five FR~I radio galaxies and isolate their nuclear
emission from that of the host galaxy. This enable us to determine the
Spectral Energy Distributions (SED) of their nuclei spanning from the
radio to the X--ray band. They cannot be described as single
power--laws but require the presence of an emission peak located
between the IR and soft X-ray band. We found consistency between the
SED peak position and the values of the broad band spectral indices of
radio galaxies with those of BL Lac, once the effects of beaming are
properly taken into account. FR~I SED are thus qualitatively similar
to those of BL Lacs supporting the identification of FR~I sources as
their mis-oriented counterparts.  No dependence of the shape of the
SED on the FR~I orientation is found.

\end{abstract}


\section{Introduction} 

In the framework of the unifying models for AGN, the low luminosity
Fanaroff--Riley (1974) type I, FR~I, radio galaxies are identified as
the parent population of BL Lac objects: the non--thermal continuum   
of BL Lacs is dominated by beaming effects, resulting from
the relativistic motion of plasma at a small angle with respect to
the line of sight (Blandford \& Rees 1978), and FR~I radio galaxies
would represent their mis-oriented counterparts.  Support to this   
unification comes from different and independent observational  
pieces of evidence (see Urry \& Padovani 1995 for a review).

In this scenario the detection of any non--thermal jet emission       
in FR~I would allow a quantitative analysis of the
unified model based on the direct comparison of the beamed component 
in the two populations.  However, to pursue this goal we need to
isolate the jet emission in the cores of radio galaxies.

Recently, Chiaberge, Capetti \& Celotti (1999) have analyzed HST
images of a complete sample of 33 FR~I radio galaxies and found an
unresolved central optical source in the majority of them.  The
luminosity of these sources show a striking linear correlation with
that of the radio cores, strongly arguing for a non--thermal
synchrotron origin also for the optical emission.  Furthermore, the   
high detection rate suggests that obscuring tori in the nuclei of
FR~I must be either geometrically thin or present only in a minority
of the objects.

HST images also reveal the presence of extended nuclear disks in low
luminosity radio galaxies (e.g. Jaffe et al. 1993).  As shown by
Capetti \& Celotti (1999, hereafter CC99) from the analysis of 5
objects, such disks can be used as reasonably good indicators of the
orientation of the radio sources. All of the 5 FR~I also show the
optical central source. By comparing their intensity with the emission
of BL Lacs with similar extended (isotropic) characteristics they
found that: i) BL Lacs are $10^{2}$--$10^{5}$ brighter than the
corresponding radio galaxies and ii) the relative core brightness
decreases with increasing observing angle, quantitatively consistent
with the inference that also the FR~I nuclear component is dominated
by the beamed jet emission.

In this framework, multi--wavelengths observations of FR~I radio
galaxies clearly represent an alternative and more detailed way to
test the model, as their SED can be directly compared with those of BL
Lacs.  The latter ones are well described by two broad peaks (in $\nu
L_{\nu}$), the first one attributed to non--thermal synchrotron
emission while the second one is most commonly ascribed to inverse
Compton scattering.  Moreover, the whole BL Lac (and Flat Spectrum
Radio Quasar, FSRQ) broad band phenomenology seems to be accounted for
by the (continuous) distribution in the energy position of the two
peaks and their relative luminosity (Padovani \& Giommi 1995; Fossati
et al. 1998): an increase in the inverse Compton vs synchrotron
luminosity ratio corresponds to a decrease in the peak frequencies, 
describing the progressive trend from FSRQ, to High and Low Energy
Peaked BL Lacs (HBL and LBL, Giommi \& Padovani 1994). This
establishes a direct relation between the type of SED/object and the
relative contribution of the two components in the X--ray band (steep
synchrotron turnover vs flat Compton power law; Padovani \& Giommi
1996).  These phenomenological trends seem also to be simply connected
with the source (bolometric and radio) power, increasing from HBL to
LBL (Ghisellini et al. 1998).

But how does orientation affect the SED ? While it has been rather
convincingly shown that orientation alone cannot account for the
differences in the SED of blazars (Sambruna, Maraschi \& Urry 1996),
the alternative interpretation just described has been - so far -
focused only on highly beamed sources (blazars), neglecting the
(comparatively smaller) effects due to orientation.

SED of radio-galaxies and in particular of FR~I have been derived by
several authors (e.g. Impey \& Gregorini 1993; Ho 1999). However, the
comparison with those of BL Lacs is hampered by the large aperture
from which data have been obtained, which often includes a dominant
contribution from the host galaxy.  We therefore gathered archival
observations of FR~I from which it is possible to separate the nuclear
from the extended emission, in particular ROSAT data and HST/NICMOS
infrared images. These can be combined with our analysis of the HST
optical images and radio core measurements from the
literature. Clearly, data in the infrared/optical and X--ray bands
could also provide an independent test for the role of absorption from
thick material (torus) in this unifying scheme.

As the orientation of the radio galaxies is a crucial parameter, in
this paper we focus on the sample of 5 objects studied in CC99 (see
Table 1), for which we have such information.

ROSAT archival pointed observations of 4 out of 5 galaxies are
available and are presented in Section 2.  HST/NICMOS infrared images
have been obtained for 3 of the selected objects and are described in
Section 3. In Section 4 we discuss the role of absorption with respect
to the inferred nuclear SED, which we assemble in Section 5. The
comparison with the SED of BL Lacs and the implications for the
FR~I/BL Lac unified scheme are discussed in Section 6, while in the
final Section 7 we present our conclusions and future perspectives.
  
Throughout this work, we adopt the following values for the
cosmological parameters: $H_{\rm 0} = 75$ km s$^{-1}$ Mpc$^{-1}$ and
$q_{\rm 0} = 0.5$.

\section{The ROSAT observations} 

ROSAT HRI and PSPC data are available for 3 of the selected sources,
namely 3C~31, 3C~270 and 3C~465, while for 3C~264 only PSPC data
exist. In Table 2 we list the available pointed observations. The
X--ray properties of these targets have been already discussed by
different authors (see last column of the table) but we decided to
re-analyze the data to follow a more homogeneous approach.

For the analysis we have used the EXSAS package (release of Jan.
1996; Zimmermann et al. 1995) and the data have been corrected for
vignetting and dead time.  In the PSPC observations only the energy
range 0.15 - 2.1 keV has been considered and the spectra have been
rebinned according to the source count rate \footnote{The minimum
signal to noise ratio per energy channel is $5 \sigma$ for 3C~264 and
3C~270, $4 \sigma$ for 3C~31 and $3 \sigma$ for 3C~465.}.
 
All targets are members of clusters or groups of galaxies, and this
implies that any central non--thermal X--ray emission must be
disentangled from that of a hot intergalactic medium, with
temperatures $\sim 3 - 8$ keV (Sarazin 1986), and of the extended hot
halos ($\sim 1$ keV; Donnelly et al. 1990; Trussoni et al. 1997).
Therefore for the spectral analysis of the PSPC data we used composite
models allowing for the presence of both a thermal and a non--thermal
(power-law) component.  The free physical parameters in the spectral
fits are, besides the two normalizations, the hydrogen column density
$N_H$, the photon index $\Gamma$ for the power--law spectrum, the
temperature $T$ and metallicity $\mu$ for the thermal component
(optically thin plasma, Raymond \& Smith 1977).  As count rates are
usually quite low, some parameters must be fixed in order to deduce
usefully constrained fits. In all cases (except for 3C~264) we set
$N_H$ to its galactic value $N_{H, gal}$, and $\mu =0.5$, as typical
for hot galactic halos (Fabbiano 1989). Fixing the metallicity does
not significantly affect our results as the parameters describing the
non-thermal component depend only weakly on $\mu$ -- at least if it
lies within the plausible range $0.2 \leq \mu \leq 1$.  To establish
the statistical significance of the non thermal component, we have
then checked through an F-test when its inclusion improves the fit.

The HRI data should allow in principle to spatially separate the
unresolved central component from the coronal emission.  However, the
spatial resolution of the HRI is insufficient to fully resolve the
cores of the galactic coronae, particularly for the more distant
sources. Furthermore, problems in the aspect solution associated with
the satellite wobble may lead to smearing of point-like sources
(algorithms proposed to overcome this problem work only for bright
sources, Morse 1994; Harris et al. 1998; Sarazin et al. 1999)
compromising the spatial analysis.  We therefore preferred to use the
HRI observations just to check their consistency with the count rates
predicted by the PSPC spectral decomposition.

No statistically significant variability has been found in the light
curves of the observations, and in the targets observed with PSPC and
HRI the count rates are consistent in both instruments. This is also
true for two very short HRI exposures of 3C~31 (Jan. 1992; see
Trussoni et al. 1997) and 3C~270 (Jul. 1994), that we have not
included in the present analysis.

\subsection{Individual sources}

In the following we discuss the X--ray properties of the targets and
in particular the results from the spectral fits with a two components
model (thermal + power-law). The results are summarized in Table 3.

\noindent
{\bf 3C~31}.  The host galaxy of 3C~31 (NGC 383) is the dominant
member of the group Arp 331. A fit to its PSPC spectrum including a
non-thermal source is more acceptable at $\approx 90$\% confidence
level than a single thermal component. We obtain a gas temperature of
$T \approx 0.6$ keV, but the spectral index of the non--thermal
component is essentially unconstrained, $\Gamma = 1.8^{+0.6}_{-1.1}$.
We derive the power-law flux adopting two values of the photon index
(namely $\Gamma=1.7$ and $\Gamma=2.3$ see Table 3): the resulting
estimates differ by only 20 \% indicating that in this case
uncertainties are dominated by photon statistics.  Our spectral
results are basically consistent with those of Komossa \& B\"ohringer
(1999).

The HRI brightness radial profile (Fig. 1a) shows a compact component
(broader than the PSF) and faint extended emission out to $40^{\prime
\prime}$.  From the PSPC analysis we expect a HRI photon flux of
$\approx 4.1 \times 10^{-3}$ cts s$^{-1}$ and $\approx 2.9 \times
10^{-3}$ cts s$^{-1}$ from the thermal and non--thermal components,
respectively.  The total HRI count rate (within the same extraction
radius of the PSPC spectrum) is $(6.6 \pm 1.3)\times 10^{-3}$ cts
s$^{-1}$, in agreement with our estimates. Furthermore, the predicted
non--thermal photon flux cannot exceed that measured in the central
compact component. Indeed within $15 \arcsec$ we measure $(5.2 \pm
0.5)\times 10^{-3}$ cts s$^{-1}$.
 
\noindent
{\bf 3C~264}. This radio galaxy lies at a projected distance of
$\approx 8^\prime$ from the center of the cluster A1367. Low
brightness emission from the cluster gas is detected around the
galaxy, but it is effectively removed after background subtraction.
The spectrum is remarkably well fitted with a single power law
($\Gamma \approx 2.46$) absorbed by a column density $N_{\rm H} =
2.6^{+0.4}_{-0.2} \times 10^{20}{\rm cm}^{-2}$, consistent with the
Galactic one ($2.55 \times 10^{20}{\rm cm}^{-2}$), in agreement with
the results of Edge \& R\"ottering (1995).  A thermal component might
be present at $\approx 1 \sigma$ level (assuming $T=1$ keV) at a
confidence level of $\approx 80$\%.

\noindent
{\bf 3C~270}. This source is associated to NGC~4261, main member of a
group of galaxies located behind the Virgo cluster. The emitted flux
is fitted better by a composite spectrum than by a single thermal law
(improving of the quality of the fit at a significance level of
$\approx 99.2$\%).  The two component spectral model provides
acceptable fits to the PSPC data with $\Gamma \approx 1.7$ and $T
\approx 0.8$ keV.  These results, and in particular the presence of a
flat non--thermal component, are confirmed by the analysis of the ASCA
observations (Sambruna et al. 1999).

The HRI brightness profile (Fig.~1b) is similar to that of 3C~31 but
the extended emission (detected up to a radius $\approx 1^{\prime}$)
is much more prominent. Again the central compact region is more
extended than the PSF.  In the HRI we detect a total count rate $(3.2
\pm 0.3) \times 10^{-2}$ cts s$^{-1}$, as expected from the PSPC data
which predict $2.2 \times 10^{-2}$ cts s$^{-1}$ and $1.1 \times
10^{-2}$ cts s$^{-1}$ for the thermal and power--law components,
respectively. The photon flux of the compact component is $(1.9 \pm
0.1) \times 10^{-2}$ cts s$^{-1}$ again consistent with the derived
intensity of the power--law one.

\noindent
{\bf 3C~465}. The galaxy NGC~7720 (3C~465) is the brightest member of
the cluster A2634. The PSPC emission from the central region is well
fitted by a thermal spectrum with $T \approx 1.2$ keV.  The addition
of a non--thermal component does not improve the fit.  We set an upper
limit to the non--thermal luminosity of $\lesssim 2 \times 10^{41}$
erg s$^{-1}$ (see Table 3).  This contrasts with the results obtained
by Hardcastle \& Worrall (1999) based on the spatial analysis of the
HRI brightness profile.

\noindent
{\bf NGC~7052}. No pointed observation is available for this source.
From the RASS an upper limit of the count rate $\lesssim 0.01$ cts
s$^{-1}$ is found (Brinkmann, private comm.). Assuming a power-law
spectrum with $\Gamma=2.3$ this corresponds to a total luminosity
$\lesssim 1.8 \times 10^{41}$ erg s$^{-1}$.
 
\subsection{Summary of the X-ray data}
Non--thermal emission is detected in 3 out of the 4 targets with
pointed X--ray observations: in 3C~264 it is the dominant source,
while in 3C~270 and 3C~31 the inclusion of a non--thermal component
significantly improves the fit.  Only an upper limit can be set for
the power-law emission from the nucleus of 3C~465.  The photon fluxes
predicted by the spectral decomposition agree with those observed in
the HRI images.

Worral \& Birkinshaw (1994) \& Worral (1997) found a correlation
between the luminosities of FR~I radio galaxies in the radio and soft
X-ray bands, recently confirmed by Canosa et al. (1999) and Hardcastle
and Worrall (1999) who analyzed the PSPC and HRI brightness profiles
of a large sample of objects.  This trend appears to hold also at
harder X-ray energies for brighter radio galaxies (Trussoni et
al. 1999a).  The non-thermal luminosity of the three detected sources
we found is consistent with this correlation.  Conversely, the upper
limit for 3C~465 is $\approx 4$ times lower than expected. Concerning
the thermal component, elliptical galaxies show a correlation between
the coronal X--ray luminosity and the optical magnitude in the $B$
band (Fabbiano et al. 1984, Donnelly et al. 1990, Eskridge et al.
1995, Trussoni et al. 1997, Beuing et al. 1999).  The thermal
luminosities derived for 3C~31 and 3C~270 (as well as the upper limit
deduced for 3C~264) are again consistent with this X-ray/optical
relation.  Only the luminosity of 3C~465 is significantly lower (by a
factor $\approx 5$) than expected.

This overall agreement provides independent support to the validity of
our spectral decomposition.

\section{The HST/NICMOS observations} 

HST/NICMOS observations are available in the public archive for
3C~264, 3C~270 and NGC~7052. For the first two objects we analyzed
images taken with the F110W, F160W and F205W wide band filters
(approximatively corresponding to the H, J and K bands, respectively),
while for NGC~7052 only F160W images exist.

The targets were observed using NICMOS Camera 2 (0\farcs075/pixel).
The field of view of this 256 $\times$ 256 pixels camera is 19\farcs4
$\times$ 19\farcs4.  All observations were carried out with a
MULTIACCUM sequence (MacKenty et al. 1997), i.e. the detector is read
out non-destructively several times during each integration to
facilitate removal of cosmic rays and saturated pixels.  The data were
re-calibrated using the pipeline software CALNICA v3.0 (Bushouse et
al. 1997) and the best reference files in the Hubble Data Archive to
produce flux calibrated images. Bad pixels were removed interpolating
from values of neighboring pixels.  A log of the observations is given
in Table 4.

The final images are presented in Fig.~2. While in NGC~7052 the
extended dusty disc seen in the optical images is clearly visible, we
do not find evidence for circum--nuclear structures in 3C~264 and
3C~270.  This is somewhat expected at this longer wavelengths where
dust absorption is less efficient, confirming that these discs are
associated with dust with small optical depth (e.g. Jaffe et
al. 1993).  On the other hand, similarly to what found in the optical
band, in all targets it is detected a central IR source, whose
emission can be now easily isolated from that of the host galaxy. We
performed aperture photometry to measure their nuclear fluxes in the
different bands (reported in Table 4).  Photometric errors are
dominated by the accuracy of the NICMOS internal flux calibration,
which is within 5\%, except in the case of NGC~7052 where the relative
error is $\sim 30 $ \%.

\section{The role of absorption in FR~I nuclei}

As absorption could have significantly affected our measurements in
the optical and X--ray bands, and thus in principle alter the inferred
SED properties, in this section we carefully examine its possible
role.

Photoelectric absorption can be very effective on soft X-rays, and
thus measures in this band provide a powerful tool to estimate the
(equivalent) column density in neutral hydrogen $N_{\rm H}$.  In
3C~264 the X-ray emission is dominated by a power--law component over
the whole ROSAT energy range and its spectrum is sufficiently well
determined that we could directly measure a column density which -- as
already mentioned -- is fully consistent with the Galactic one: no
significant local absorption is present in this source.  This is
somewhat expected as 3C~264 is likely to be seen at a small angle from
the jet axis (see Table 1) and thus along an unobscured line of sight
(within the nuclear region).  Conversely, in 3C~31, 3C~270 and 3C~465,
the observed X-ray flux is dominated by extended thermal emission
which is unaffected by nuclear absorption. The detection in 3C~31 and
3C~270 of non--thermal emission above 1 keV only sets an upper limit
of $N_{\rm H} \lesssim 10^{23}$ cm$^{-2}$\footnote{This column density
corresponds to an attenuation of a factor 100 in the intensity at 2
keV}.  Note that these results are in agreement with the conclusions
of Chiaberge et al. (1999), namely that optically thick material
(torii) affects the nuclear emission at most in a minority of FR~I.
Clearly only X-ray observations at higher spatial resolution,
increasing the fraction of nuclear vs extended component, can provide
more stringent and valuable estimates of $N_H$ in these cases.

Obviously optical images do not directly lead to estimates of
absorption. However, the tight linear correlation which exists between
the optical (R band) and radio fluxes/luminosities of FR~I cores (see
Introduction) can set interesting upper limits on the nuclear
extinction. In fact, even if the entire dispersion of this correlation
(0.4 orders of magnitude) is due unically to foreground dust
absorption varying from source to source, the extinction can not
exceed $\sim$ 2 magnitudes.  This limit translates into a gas column
density of $N_{\rm H}\lesssim 5 \times 10^{21}{\rm cm}^{-2}$ (assuming a
standard gas-to-dust ratio and the extinction curve by Cardelli,
Clayton \& Mathis 1988), and thus the flux at 1 keV could be depressed
at most by a factor of 4 (photoelectric absorption values tabulated by
Morrison \& McCammon 1983).

Given these upper limits on the exctintion, what would be its effect
on the SED properties (and more specifically on the broad band
spectral indeces)?  As the fluxes in the V band and at 1 keV can be
underestimated by a factor $\lesssim 10$ and $\lesssim 4$
respectively, the corresponding (maximum) changes in the spectral
indeces are limited to $\Delta \alpha_{\rm RO} \lesssim -0.20$,
$\Delta \alpha_{\rm RX}\lesssim -0.08$ and $\Delta \alpha_{\rm OX}
\lesssim 0.15 $ ($F_{\nu} \propto \nu^{-\alpha}$ \footnote{they are
calculated between 5 GHz, 5500 \AA and 1 keV (rest frame)}). Such
uncertainties do not affect significantly our findings and
conclusions. 

\section{Spectral Energy Distributions of FR~I nuclei}

Combining the available data in different energy bands, it is possible
to build a SED for the FR~I nuclei, isolating for the first time this
emission from that of the host galaxy (see Fig.~3 and Table 5).
Besides the IR and X--ray data presented in the previous sections, we
used HST optical fluxes from CC99 and radio data taken from the
literature. In the following we describe the individual SED -- always
referring to a $\nu L_{\nu}$ representation -- and characterize them
through the broad band spectral indices, $\alpha_{\rm RO}$,
$\alpha_{\rm RX}$ and $\alpha_{\rm OX}$ (the potential effects
produced by the presence of local absorption are discussed in the
previous Section 4).

\noindent
$\bullet$ {\bf 3C~31}. The average radio-to-optical spectral index is
$\alpha_{\rm RO} = 0.79$ while the optical-to-X--ray one is
$\alpha_{\rm OX} = 1.03$.  These spectral indices imply the presence
of a peak in the energy distribution. A better energy coverage or the
determination of the optical (or X--ray) spectral slope are necessary
in order to better constraint its location.

\noindent
$\bullet$ {\bf 3C~264}. Its SED is the better determined in our sample
as there are data at radio, infrared and optical frequencies, the
X--ray spectrum is also well defined and the absorption is constrained
to a negligible value.  The energy distribution rises from the radio
toward higher frequencies (with an average index $\alpha_{\rm RO} =
0.63$) and it is still rising in the IR/optical band, although with a
flatter slope.  At 1 keV the intensity is 20\% larger than in the
optical (in $\nu$L$_{\nu}$) but the X--ray spectrum is rather steep
($\alpha_{\rm X}$= 1.45) indicating that the SED has already reached
its peak, which must be thus located in the interval $10^{15} -
10^{16.5}$ Hz.

\noindent
$\bullet$ {\bf 3C~270}. This SED is characterized by the sharp decline
in the IR/optical range described by a spectral index $\alpha \sim
3.5$.  Clearly dust absorption can play a significant role in
producing this rapid decline.  However its spectrum in the X--ray band, as
also measured by ASCA, indicates that the minimum observed at
optical energies is real, although possibly enhanced by the presence
of dust. Moreover, its representative point is not located
significantly below the radio-optical correlation described by
Chiaberge et al. (1999), arguing against an obscuration larger than
average.

\noindent
$\bullet$ {\bf 3C~465}. After rising from the radio to the optical
($\alpha_{\rm RO} = 0.83$), the SED of 3C~465 is already falling at
higher energies as the X--ray upper limit lies below the optical
point. This indicates that also in this case a peak in the emission is
reached below the X--ray band, although the poorly determined SED does
not allow us to better constrain its location.

\noindent
$\bullet$ {\bf NGC~7052}. The average radio-to-optical spectral index
is $\alpha_{\rm RO} = 0.80$. Contrarely to what seen in 3C~270, the
nuclear source in NGC~7052 is (marginally) fainter in the infrared
than in the optical band. This indicates that, although the extended
disk is seen at the highest inclination ($\sim$ 70$^\circ$) and has a
very large optical depth, the nuclear emission is not significantly
affected by dust obscuration.  The upper limit derived from the RASS
requires a flattening of the SED slope at higher energies.

Summarizing, the nuclear SED of the selected FR~I present a wide
variety of behaviors.  However three common features can be
recognized: i) it always rises from the radio to the infrared/optical;
ii) no SED can be described as a single power law; iii) for all of the
sources with pointed X--ray data the presence of a peak in the
emission can be localized at energies lower than soft X--rays.

\section{Discussion}

\subsection{Comparison of the SED of FR~I and BL~Lacs}

Let us now compare the SED of FR~I with those of BL Lacs.  Despite of
the poor spectral coverage of the FR~I SED (somewhat reminiscent of
the SED coverage of BL Lac objects in the 80s), it is possible to
recognize a qualitative similarity with the SED of blazars, and in
particular the likely presence of one broad peak (see Fig.~3). In
fact, as discussed in Section 5, the spectra appear to raise from the
radio to the IR/optical band, and in all cases - except for NGC~7052 -
the X--ray information (flux and/or spectrum) suggest that a peak in
the SED occurs at lower energies.

In the specific case of 3C~270 the flat X--ray spectrum is plausibly
due to the dominance of a rising Compton component in this band.  And
interestingly its extremely steep IR-optical spectrum seems to
indicate the presence of a minimum, fully consistent with the
indications derived from the X--ray spectral shape.  Agreement between
the hints inferred from the IR/optical and X--ray spectra is also
found in the SED of 3C~264 in which the X--rays can be interpreted as
the steep high energy cut-off of the synchrotron component.

Clearly, the paucity of spectral information limits any detailed
analysis and in particular does not allow us to determine the location
of the energy peak precisely enough to identify the 'type' of blazar
putatively hosted in the radio galaxies, except possibly for 3C~270
and 3C~264, which resemble an LBL and HBL, respectively.

Nevertheless, it is possible to profitably use the available data to
compare the broad band spectral indices with those of BL Lacs, as
shown in Fig.~4 where complete samples of BL Lacs (and FSRQ) are
considered
\footnote{The separation between LBL and HBL does not univocally
correspond to the selection band of BL Lacs (i.e. Radio selected BL
Lacs and X-ray selected BL Lacs, RBL and XBL).  However, there is a
sufficiently large overlap of RBL with LBL (and of XBL with HBL) that
we consider the values of spectral indices for RBL and XBL as
representative of the respective classes as defined from the SED peak
location.}.  Three out of the four radio galaxies (3C~31, 3C~264 and
3C~465) lie in the area occupied by FSRQ and RBL.  Particularly
puzzling is the case of 3C~264. In fact this source is observed at a
small angle with respect to the jet axis, re-enforcing the view that
its nucleus should be rather similar to that of a BL Lac. Furthermore
its SED suggests that the putative synchrotron peak is located between
the optical and X--ray bands, analogously to HBL, in contrast with
its position in the spectral index planes. The fourth source, 3C~270,
is located on its own in the upper left corner of the diagrams.

However, as discussed in detail by Chiaberge et al. (2000)
relativistic beaming can have strong effects on the position of the
source in the $\alpha_{\rm ro}$--$\alpha_{\rm ox}$ plane.  Due to the
shift in energy caused by beaming, these quantitatively depend on the
shape of the SED if there are changes in the spectral slope (even in
the simplest case in which the degree of beaming does not depend on
energy).  The transformation law for broad band spectral indices,
derived in Chiaberge et al. (2000), is \\
\begin{displaymath}
\alpha_{\rm BL Lac} - \alpha_{\rm FR~I} = (\alpha_1-\alpha_2)
	\frac{\log (\delta_{\rm BL Lac}/\delta_{\rm FR~I})} 
                    {\log (\nu_2/\nu_1)} \, ,
\end{displaymath}
where $\alpha_1$, $\alpha_2$ are the radio, optical or X-ray spectral
indices and $\delta_{\rm BL Lac}$, $\delta_{\rm FR~I}$ are the
relativistic beaming factors.

For 3C~264 and 3C~270, for which $\alpha$ in both the optical
and X-ray bands are measured,
we can determine ``beaming tracks'', i.e. the changes of the broad
band spectral indices for increasing Doppler boosting (Fig.~4),
adopting $\alpha_r = 0$. For
3C~264, since the SED slope in the X--ray is steeper than in the
IR/optical and radio bands, blue-shifting the SED has the effect of
reducing all spectral indices. Its representative point thus moves
toward the origin in both the $\alpha_{\rm ox}$--$\alpha_{\rm ro}$ and
$\alpha_{\rm ox}$--$\alpha_{\rm rx}$ planes and the beaming track
crosses the region in which HBL are located.  Similarly, the track of
3C~270 intercepts the region typical of LBL
\footnote{As dust absorption can cause an enhancement of the minimum
in the IR/optical region of the SED of 3C~270, we also estimated its
`beaming track' allowing for the presence of extinction. This turns
out to be similar to the case of no absorption.}. For both objects, the
inconsistency between SED shape and the values of the spectral indices
might be thus resolved, once the effects of beaming are properly taken
into account.

We note, however, that the beaming factor required in the case of
3C~264 is ${\delta_{\rm BL Lac}}/{\delta_{\rm FR~I}} \sim 10 - 100 $
(the beaming correction for 3C~270 is heavily dependent on possible
effects of dust extinction and thus is not quantitatively reliable).
As the nuclear luminosity increases as $\delta^{p+\alpha}$ (with $p= 2
- 3$), when seen pole-on 3C~264 would be 3 - 6 orders of magnitude
brighter in the optical band (where $\alpha \sim 1$ and for $p = 2$).
This is only marginally consistent with the estimates of the ratio
between its luminosity and that of BL Lacs of similar extended
(isotropic) power, which are only $\sim 10^2 - 10^3$ times brighter
(CC99).  Furthermore, as discussed in the Introduction
the radio luminosity of BL Lacs appears to be
univocally connected with their SED. As the nuclear radio luminosity
of 3C~264 is $\nu L_{\nu} \sim 10^{39.9}$ erg s$^{-1}$, if seen
pole-on this would increase by $\sim 10^2 - 10^4$ (again for $p = 2$)
reaching values which are typical of LBL and higher than those seen in
HBL (Fossati et al. 1998).  Chiaberge et al. (2000) found a similar
problem when comparing the luminosities of complete samples of FR~I
and BL Lacs: FR~I appear to be overluminous with respect to what is
expected by debeaming the BL Lacs emission, modeled with a single
emitting component.  They suggested that this can be ascribed to a
velocity structure within the relativistic jet and that in FR~I the
emission is dominated by components/regions moving slower than those
seen in BL Lacs.

Summarizing, the SED of FR~I are qualitatively similar to those of BL
Lac objects, supporting the identification of these sources as their
mis-oriented counterparts, but their intensity does not seem to be
compatible with the simplest model in which we are observing a single
beamed emitting component in both classes of objects.

\subsection{X-ray emission and radio-galaxy orientation}

Following the method developed in CC99, we compare the luminosity of
FR~I nuclei with that of BL Lac objects in the X-ray band. In order to
restrict the comparison to objects which potentially belong to the
same region of the luminosity function of the FR~I/BL Lacs population,
we identify BL Lacs whose isotropic properties are similar to those of
our radio-galaxies.  More specifically, we consider objects whose
extended radio luminosity (data taken from Kollgaard et al. 1996)
differs by less than a factor of two from each of our FR~I.
\footnote{ With respect to CC99, we dropped the further requirement of
similar host galaxy magnitude since this does not appear to strongly
correlate with the nuclear properties neither in BL Lacs nor in
FR~I. This less restrictive choice enables us to increase the number
of BL Lac counterparts for each FR~I.}  We thus selected between 6 and
10 `relatives' for each FR~I. Their X--ray fluxes (1 keV rest frame
energy, corrected for absorption) and spectral indices in the ROSAT
band are taken from Lamer et al. (1996) and Urry et al. (1996).  The
X-ray nuclear emission of FR~I is fainter than that of BL Lacs by a
factor which ranges from $\sim$ 10 up to 3$\times$10$^4$.

For a given FR~I, the X-ray luminosity ratios with respect to the
different `relative' BL Lacs present a very large dispersion, of 2 --
3 orders of magnitude, and therefore, contrarely to what is found in
the optical band (CC99) we cannot draw any conclusion on whether a
trend with orientation is present.

\section{Summary and future perspectives} 

We used ROSAT and HST observations to isolate the emission originating
from the nuclei of 5 FR~I sources with the aim of studying their
Spectral Energy Distributions.  For the ROSAT PSPC observations, in
order to disentangle the nuclear non-thermal emission from that of the
host galaxy halo, we performed fits allowing for the presence of both
a thermal and a power--law component: in the case of 3C~264 the
emission is dominated by a power law spectrum; in 3C~31 and 3C~270 the
inclusion of the non--thermal component improves the fit, while this
is not required in 3C~465. A nuclear source is clearly seen in the
HST/NICMOS infrared images, available for 3 objects.

Combining the measurements in the X-ray and IR band with optical
nuclear luminosities derived from HST/WFPC2 images and radio data
taken from the literature, we have been able to build for the first
time the SED of FR~I nuclei.

These indicate the presence of a peak between the IR and the soft
X--ray bands, and the X--ray slopes are consistent with the broad band
indications.  For the two best constrained SED, it is also possible to
tentatively classify the FR~I nuclei similarly to what is done for BL
Lacs, in High and Low Energy Peaked sources.  This turns out to agree
with that derived from the comparison of the broad band spectral
indices once the changes in the SED due to the beaming are properly
taken into account.  However, the intensity of the nucleus of 3C~264
does not appear to be compatible with the simplest model in which we
are observing a single beamed emitting component in both FR I and BL
Lacs.  The SED of FR~I are thus qualitatively (but probably not
quantitatively) similar to those of BL Lacs.

The proposed approach of comparing the SED of the parent and beamed
population appears to be promising. Better data and energy coverage
are clearly required, particularly in the X--ray band, as the spectral
shape at these energies is essential in giving clues on the frequency
range where most of the synchrotron emission is released and thus in
`classifying' FR~I sources similarly to what is done for BL Lacs.
Both the AXAF and XMM missions will soon provide us with X--ray
spectra for large, and especially complete, samples of FR~I.

\section*{Acknowledgments} 

The authors would like to thank the referee for their useful comments
and suggestions.
The authors acknowledge the Italian MURST for partial financial support
under grant Cofin 98-02-32 and from the Italian Space Agency (ASI).
This research was supported in part by the National Science Foundation
under Grant No. PHY94-07194 (A. Celotti). We wish to thank I. Lehmann for 
providing the inflight calibrated PSF of the HRI, and W. Brinkmann for
allowing us to access to the RASS data on NGC 7052.

\section*{References} 

%\refitem Abraham R.G., McHardy I.M., Crawford C.S., 1991, MNRAS, 252,
%482

%\refitem Antonucci R.R.J., Ulvestad J.S., 1985, ApJ, 294, 158

\refitem Beuing J., D\"obereiner S., B\"ohringer H., Bender R., 1999, MNRAS 302, 209

\refitem Blandford R.D., Rees M.J., 1978, in BL Lac Objects,
A.N. Wolfe, ed., Univ. Pitt. Press (Pittsburgh), p.~328

\refitem Buote D.A., Fabian A.C., 1998, MNRAS 296, 977    

\refitem Bushouse H., Skinner
C.J., MacKenty J.W, 1997, NICMOS Instrument Science Report, 97-28
(Baltimore STScI)

\refitem Canosa C.M., Worrall D.M., Hardcastle M.J., Birinshaw M.,  
1999, MNRAS, 310, 30

\refitem Capetti A., Celotti A., 1999, MNRAS, 304, 434 (CC99)

\refitem 
Cardelli, J.A., Clayton, G.C., Mathis, J.S. 1988, ApJ, 329, L33 

%\refitem Celotti A., Maraschi L., Ghisellini G., Caccianiga A.,
%Maccacaro T., 1993, ApJ, 416, 118

\refitem Chiaberge, M., Capetti A., Celotti A.,
1999, A\&A 349, 77

\refitem Chiaberge, M., Celotti A., Capetti A., Ghisellini G.,
2000, A\&A 358, 104

\refitem Davis D.S., Mushotzky R.F., Mulchaey J.S., Worrall D.M., 
Birkinshaw M., Burstein D., 1995, ApJ 444, 582

%\refitem di Serego Alighieri S., Danziger I.J., Morganti R., Tadhunter
%C.N., 1994, MNRAS, 269, 998

\refitem Donnelly H.L., Faber S.M., O'Connell R.M., 1990, ApJ 354, 52 

\refitem Edge A.C., Rottgering H., 1995, MNRAS, 277, 1580

\refitem Eskridge P.B., Fabbiano G., Dong-Woo K., 1995, ApJS 97, 141

\refitem Fabbiano G., 1989, ARAA 27, 87

\refitem Fabbiano G., Miller L., Trinchieri G., Lomgair M., Elvis M., 1984, ApJ 277, 115 

%\refitem Falomo R., Urry C.M., Pesce J.E., Scarpa R., Giavalisco M.,
%Treves A., 1997, ApJ, 476, 113

\refitem Fossati G., Maraschi L., Celotti A., Comastri A., Ghisellini
G., 1998, MNRAS, 299, 433

\refitem Fanaroff B.L., Riley J.M., 1974, MNRAS, 167, 31p

%\refitem Ghisellini G., Padovani P., Celotti A., Maraschi L., 1993,
%ApJ, 407, 65

\refitem Ghisellini G., Celotti A., Fossati G., Maraschi L., Comastri, A.,
1998, MNRAS 301, 451

\refitem Giommi P., Padovani P. , 1994, MNRAS, 268, L51

\refitem Hardcastle M.J., Worrall D.M., 1999, MNRAS, 309, 969

\refitem Harris D.E., Silverman J.D., Hasinger G., Lehman I., 1998, A\&AS 133, 431

\refitem Impey C., Gregorini L., 1993, AJ, 105, 853

\refitem Jaffe W., Ford H.C., Ferrarese L., Van den Bosch F.,
O'Connell R.,W.  1993, Nat, 364, 213

\refitem Ho L.C. 1999, ApJ 516, 672 

%\refitem Kollgaard R.I., Wardle J.F.C., Roberts D.H., Gabuzda D.C.,
%1992, AJ, 104, 1687

\refitem Kollgaard R.I., Palma C., Laurent-Muehleisen 
S.A., Feigelson E.D. 1996, ApJ 465, 115 

\refitem Komossa S., B\"ohringer H., 1999, A\&A 344, 755

\refitem Lamer G., Brunner H., Staubert R., 1996, A\&A, 311, 384

\refitem MacKenty J.W., et al.,
1997, NICMOS Instrument Handbook, Version 2.0 (Baltimore STScI)

\refitem Matsumoto H., Koyama K., Awaki H., Tsuru T., Loewenstein M., Matsushita K., 1997, ApJ 482, 133

\refitem Morrison R., McCammon D., 1983, ApJ, 270, 119
   
\refitem Morse J.A., 1994, PASP 106, 675

\refitem Padovani P. , Giommi P., 1995, ApJ, 444, 567

\refitem Padovani P. , Giommi P., 1996, MNRAS, 279, 526

%\refitem Padovani P., Urry C.M., 1990, ApJ, 356, 75

%\refitem Padovani P., Urry C.M., 1991, ApJ, 368, 373

\refitem Perlman E.S. , Stocke J.T., 1993, ApJ, 406, 430

\refitem Raymond J.C., Smith B.W., 1977, ApJS 35, 419 

\refitem Sakelliou I., Merrifield M.R., 1998, MNRAS 304, 434

\refitem Sambruna R.M., Maraschi L., Urry C.M., 1996, ApJ, 463, 444

\refitem Sambruna R.M., Eracleous M., Mushotzky R.F., 1999, ApJ, 526, 60

\refitem Sarazin C.L., 1986, Rev. Mod. Phys. 58, 1

\refitem Sarazin C.L., Koekemoer A.M., Baum S.A., O'Dea C.P., 
Owen F.N., Wise M.W., 1999, ApJ 510, 90

\refitem Schindler S., Prieto M.A., 1997, A\&A 327, 37
             
%\refitem Stickel M., K\"uhr H., 1993, A\&A, 98, 39

%\refitem Stickel M., Padovani P., Urry C.M., Fried J.W., K\"uhr H.,
%1991, ApJ, 374, 431

\refitem Tananbaum H., Tucker W., Prestwich A., Remillard R., 1997, ApJ 476, 83

\refitem Trussoni E., Massaglia S., Ferrari R., Fanti R., Feretti L.,
Parma P., Brinkmann W., 1997, A\&A 327, 27

\refitem
Trussoni E., Vagnetti F., Massaglia S., Feretti L., Parma P., Morganti R., 
Fanti R. Padovani P., 1999a, A\&A 348, 437

\refitem
Trussoni E., Feretti L., Capetti A., Celotti A. Chiaberge M., 1999, Ap. Lett.
Comm., in press 

%\refitem Ulrich M-H, 1989, in BL Lac Objects: 10 years after, Maraschi
%L., Maccacaro T. \& Ulrich M.-H. eds, Springer-Verlag, p. 45

\refitem Urry C.M., Padovani P., 1995, PASP, 107, 803

\refitem Urry, C.M., et al., 1996, ApJ, 463, 424

%\refitem Urry C.M., Padovani P., Stickel M., 1991, ApJ, 382, 501

\refitem Worrall D.M., Birkinshaw M., 1994, ApJ, 427, 134

\refitem Worral D.M., 1997, in Relativistic Jets in AGN's, M. Ostrowski, 
M. Sikora, G. Madjeski, M. Begelman eds., Astron. Obs. of the Jagiellonian Univ., Krakow, p. 20

\refitem Zimmermann H.U., Becker W., Belloni T., D\"obereiner S., Izzo C., 
Kahabka P., Schwentker O., 1995, EXSAS User's GUide, ed. 5, MPE Rep. n 257

\begin{table*}
\caption{Main optical parameters}
\begin{tabular}{|l c c c c c |} \hline
Source   & Alt. name & z & $M_{B_{T_o}}$ & Envir. & Orient.$^{\rm a}$\\
\hline
3C~31      & NGC~383  & 0.0169  & -21.19   &   Arp~331  & 35$^\circ$ \\
3C~264     & NGC~3862 & 0.0216  & -21.15   &   A~1367   & 15$^\circ$ \\
3C~270     & NGC~4261 & 0.0074  & -21.02   &   Group    & 65$^\circ$ \\
3C~465     & NGC~7720 & 0.0291  & -22.30   &   A~2634   & 45$^\circ$ \\
NGC~7052   & B2 2116+26  & 0.0164  & -21.41   &   Pair  & 72$^\circ$ \\
\hline 
\end{tabular} 
\vskip 0.1 true cm 
\noindent 
$^{\rm a}$ Radio source orientation taken from CC99.
\end{table*}

\begin{table*}
\caption{Log of ROSAT observations for the four FR~I radio galaxies}
\begin{tabular}{|l c c c c c c c |} \hline
Source   & Instrument & Obs. Date &  t$_{\rm exp}$ (s) & Counts$^{\rm a}$ 
& $r^{{extr}}_{source}$ & $r^{{extr}\,{\rm (c)}}_{bkg}$  &   Ref.$^{\rm d}$ 
\\  \hline
3C~31    & PSPC  & Jul 91   & 29430 &   $552 \pm 38$  & $75^{\prime \prime}$
         &  $125^{\prime \prime}$ &   (1)(2)\\
         & HRI   & Apr 94   & 25007 &   $165 \pm 33$  &  
         &    &  (2)(3) \\
3C~264   & PSPC  & Nov 91   & 18745 &  $5068 \pm 85$  & $100^{\prime \prime}$
         &  $175^{\prime \prime}$ &   (3)(4)(5)\\
3C~270   & PSPC  & Dec 91   & 21863 &  $1935 \pm  65$  & $150^{\prime \prime}$
         &  $250^{\prime \prime}$ &   (6) \\
         & HRI   & Jun 95   & 18477 &   $584 \pm 51$   &  
         &  &      \\
3C~465   & PSPC  & Jun 91   &  9866 &      &              
         &       &       \\
         & PSPC  & Dec 92   & 10522 &   $252 \pm 21^{\rm b}$ & 
           $35^{\prime \prime}$ &  $75^{\prime \prime}$ &  (7)  \\
         & HRI   & Jan 95   & 29063 &      &
         &       &        \\
         & HRI   & Jun 95   & 33392 &   $304 \pm 29^{\rm b}$  &
             &   &  (3)(8)  \\
\\
\hline
\end{tabular}
\vskip 0.1 true cm 
\noindent 
$^{\rm a}$ Background subtracted; the errors are at $1 \sigma$

\noindent
$^{\rm b}$ Merging the data of the two observations

\noindent
$^{\rm c}$ External radius, the inner radius $\equiv
r^{extr}_{source}$

\noindent
$^{\rm d}$ References: (1) Trussoni et al. (1997); (2) Komossa \&
B\"ohringer (1999); (3) Hardcastle \& Worrall (1999); (4) Edge \& R\"ottering 
(1995); (5)Tananbaum et al. (1997);
(6) Worral \& Birkinshaw (1994); (7) Schindler \& Prieto
(1997); (8) Sakelliou \& Merrifield (1997)

\end{table*}

\begin{table*}
\caption{Fits  of PSPC data with a thermal   +  power-law spectrum.
Errors are at $1 \sigma$. Upper and lower limits are at $2 \sigma$ 
confidence level. Quantities without errors are fixed. 
Metallicity is always set to $\mu=$ 0.5 solar.}
\begin{tabular}{|l c c c c c c  |} \hline
Source   & $N_{H}$ &  $T$  & 
$\Gamma$ & $\chi^2$ (d.o.f.) & $L_{\rm [0.1-2.4 \, keV],th}$   &  
$L_{\rm [0.1-2.4 \,keV],pl}$  \\
         &$\times 10^{20}$ cm$^{-2}$& keV  & 
&    &   $\times 10^{41}$ erg s$^{-1}$ 
&  $\times 10^{41}$ erg s$^{-1}$  \\

\hline
3C~31    &  5.23 &  0.75$^{+0.17}_{-0.20}$   &    2.3 & 0.30 (15) 
         & $0.65^{+0.17}_{-0.16}$   &  $1.30^{+0.54}_{-0.58}$   \\
         &  5.23 &  0.59$^{+0.17}_{-0.20}$   &     1.7 & 0.30 (15) 
         &   $0.70^{+0.20}_{-0.22}$ &  $1.00^{+0.38}_{-0.23}$   \\
\hline
3C~264   & $2.6^{+0.4}_{-0.2}$ & 1  &  $2.46^{+0.06}_{-0.06}$ 
         & 0.88 (81) & $1.00^{+1.17}_{-1.00}$ &  $47.1^{+0.9}_{-1.3}$   \\
\hline
3C~270   & 1.63 &  $0.81^{+0.09}_{-0.13}$ &  $1.71^{+0.21}_{-0.18}$ 
         & 0.70  (33)   & $0.49^{+0.08}_{-0.06}$   &  $0.54^{+0.1}_{-0.1}$ \\
\hline
3C~465   &  5.22 &  $1.16^{+0.20}_{-0.15}$ &    2.3  & 1.10 (17)
         &  $3.20^{+0.27}_{-0.76}$ ($\gtrsim 2.2$) &  0 ($\lesssim 1.3$)  \\
         &  5.22 &  $1.16^{+0.14}_{-0.18}$ &    1.7  & 1.09 (17)
         & $3.07^{+0.36}_{-0.49}$ ($\gtrsim 1.2$)  &  0 ($\lesssim 2.1$)   \\
\hline
\end{tabular}
\vskip 0.1 true cm
\noindent
\end{table*}

\begin{table}
\caption{Log of NICMOS observations}
\begin{tabular}{|l c c c c|} \hline
Source   & Obs. Date & Filter &  t$_{\rm exp}$ (s) & 
Flux$^{\rm a}$ \\ 
\hline
3C~264     & May 98 & F110W & 112   & 6.7  \\
           &        & F160W & 112   & 4.3 \\
           &        & F205W & 112   & 2.6 \\
3C~270     & Apr 98 & F110W & 192   & 0.69 \\
           &        & F160W & 192   & 1.2 \\
           &        & F205W & 192   & 1.6 \\
NGC~7052   & Aug 98 & F160W & 160   & 0.06  \\
\hline
\end{tabular}
\vskip 0.1 true cm
$^{\rm a}$Fluxes in units of 10$^{-17}$~erg~s$^{-1}$~cm$^{-2}$~\AA$^{-1}$ 
\end{table}

\begin{table*}
\caption{FR~I nuclear luminosities$^{\rm a}$ and spectral indeces}
\begin{tabular}{@{}lccccccc}\hline
Name  & L$_{\rm R}$ & L$_{\rm IR}$ & L$_{\rm O}$ & L$_{\rm X}$ 
& $\alpha_{\rm RO}$ & $\alpha_{\rm RX}$ & $\alpha_{\rm OX}$ \\
\hline
3C~31     & 39.57  &  ---   & 40.63 &     40.49  
& 0.79 $\pm$ 0.03  &  0.88 $\pm$ 0.05 & 1.05 $\pm$ 0.12 \\ 
3C~264    & 39.94 & 41.77 & 41.81 &     42.01  
& 0.63 $\pm$ 0.02 &  0.73 $\pm$ 0.02 &  0.93 $\pm$ 0.05 \\ 
3C~270    & 38.96 & 40.30 & 39.08 &     40.32  
& 0.98 $\pm$ 0.04 &  0.82 $\pm$ 0.04 &  0.54 $\pm$ 0.09 \\ 
3C~465    & 40.34 &  ---   & 41.21 & $<$40.91   
& 0.83 $\pm$ 0.02 &  $>$ 0.92 & $>$ 1.11  \\ 
NGC~7052  & 39.07 & 39.72 & 40.04 &       --   
& 0.81 $\pm$ 0.06 &  --- &  --- \\ 
\hline
\end{tabular}
\vskip 0.1 true cm
$^{\rm a}$ 
Luminosities, log ($\nu L_{\nu}$), in erg s$^{-1}$ measured
at 5 GHz, 1.6 $\mu$m, 0.5 $\mu$m and 1 keV. 
\end{table*}

%Figure 1a:
\begin{figure*}[t]
\centerline{\psfig{figure=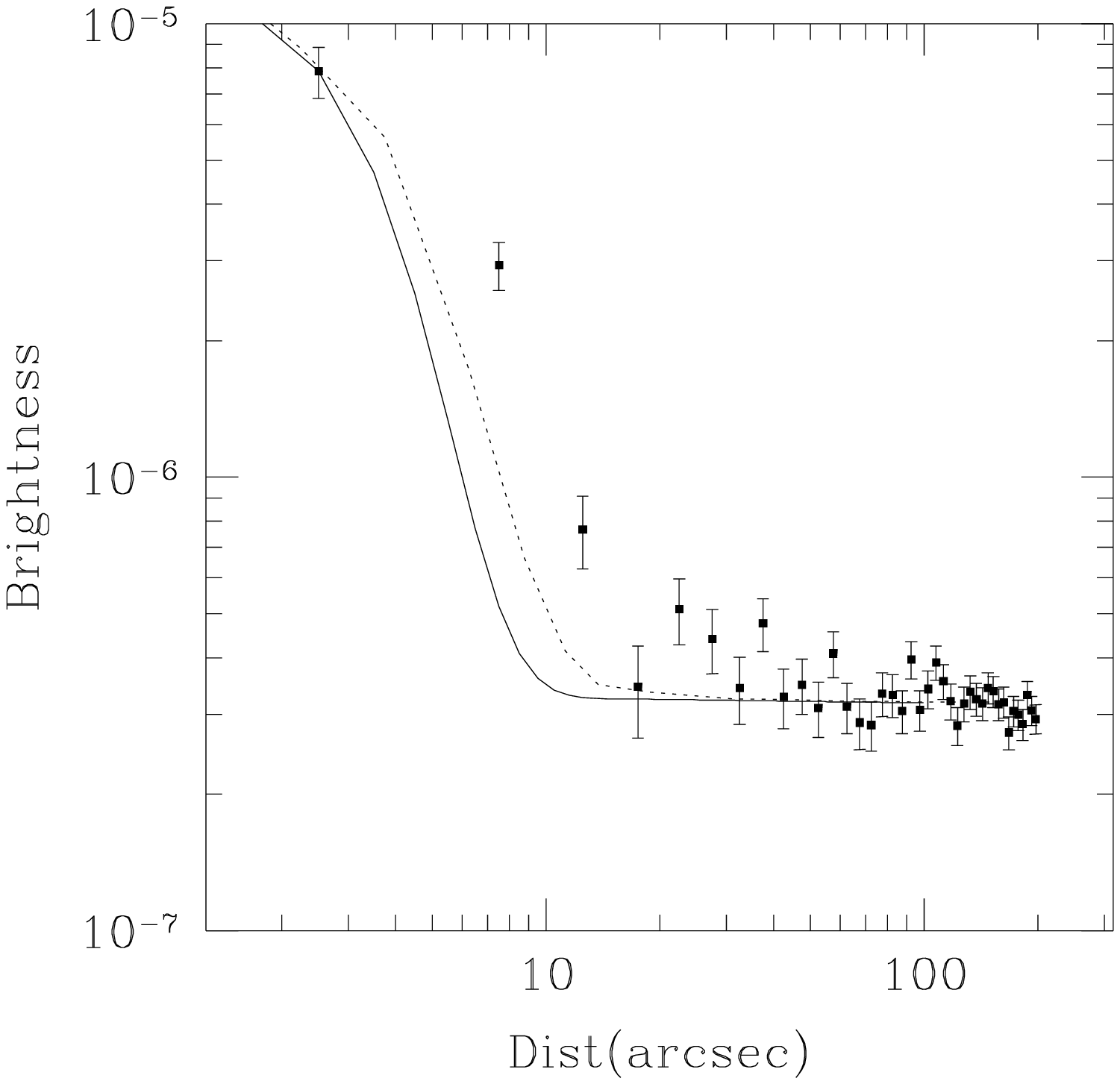,height=10.0truecm,angle=0}
\psfig{figure=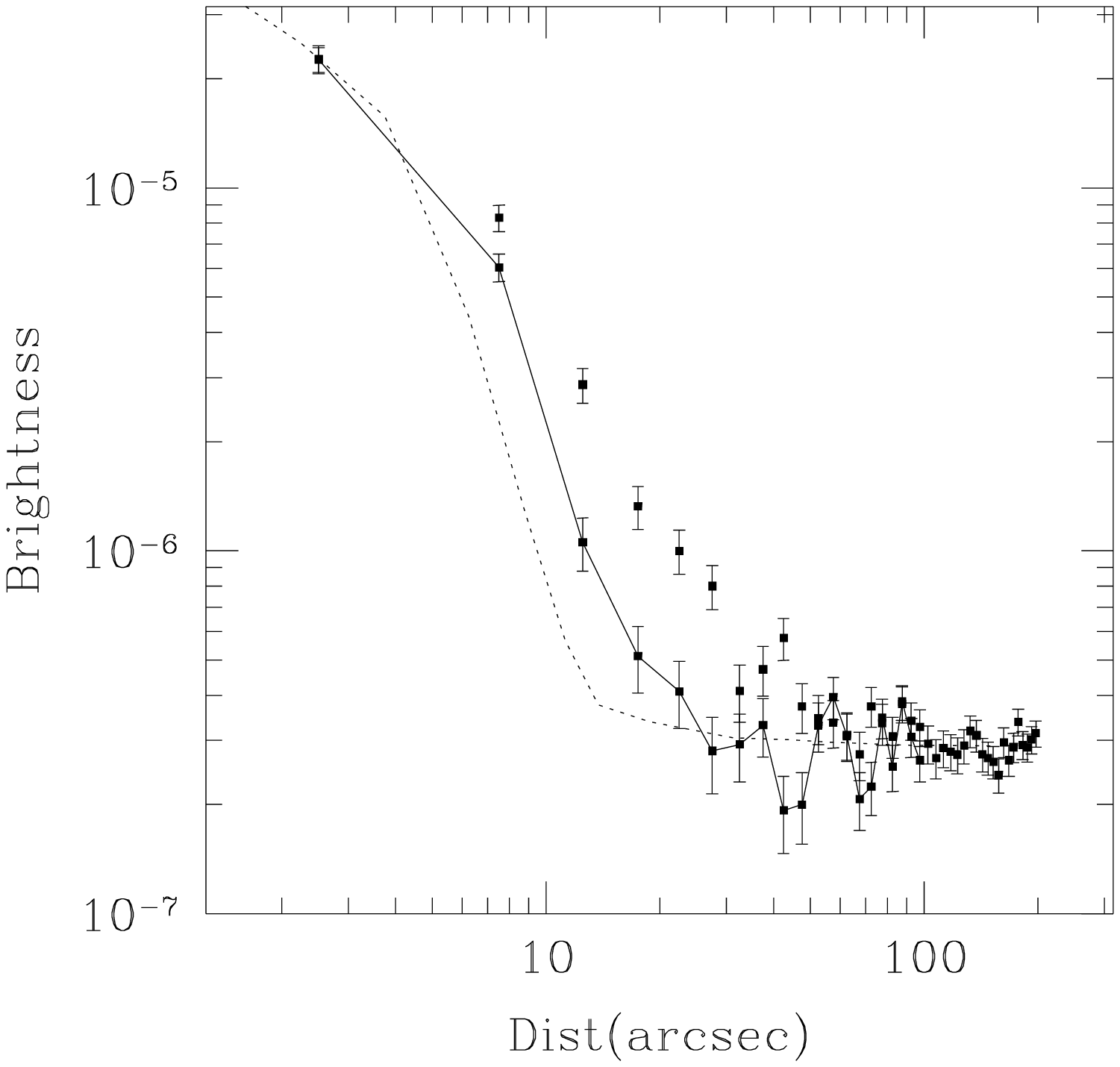,height=10.0truecm,angle=0}}
\caption{Radial profiles of 3C~31 and 3C~270 (left and right panel,
respectively), compared to the in--flight calibrated PSF (dotted line)
normalized to the source intensity and background level.}
\end{figure*}

%Figure 2
\begin{figure*}[t]
\centerline{\psfig{figure=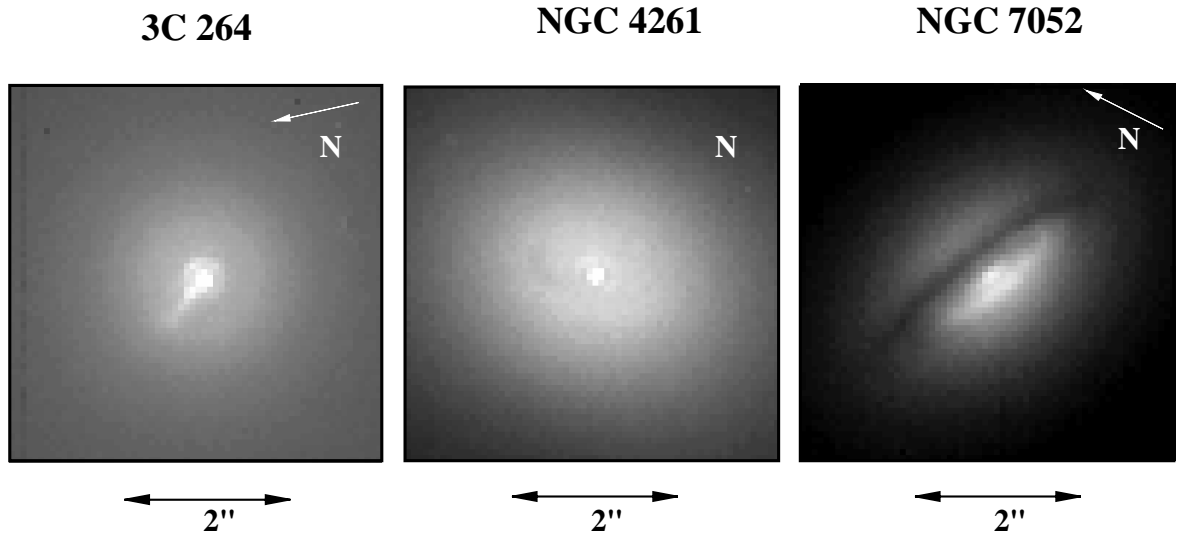,height=9.0truecm,angle=270}}
\caption{HST/NICMOS F160W images of 3 FR~I radio galaxies. Note the
presence of the central nuclear sources and of the dusty disc in
NGC~7052. The arrows indicate the direction of the radio jet.}
\end{figure*}

%Figure 3
\begin{figure}[t]
\centerline{\psfig{figure=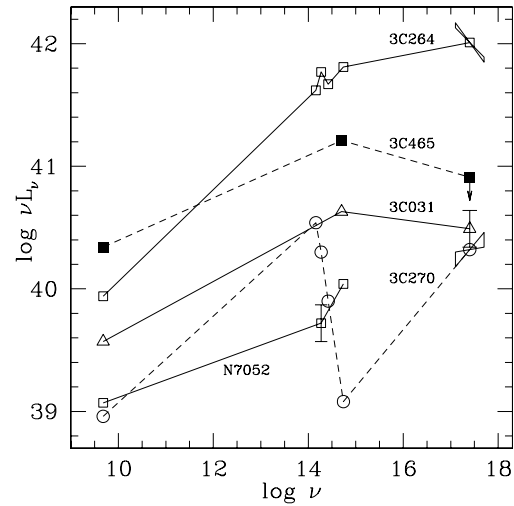,height=7.0truecm,angle=0}}
\caption{Spectral energy distribution of FR~I nuclei. 
Errors are smaller than the symbol size, except when marked.}
\end{figure}

%Figure 4
\begin{figure}[t]
\centerline{\psfig{figure=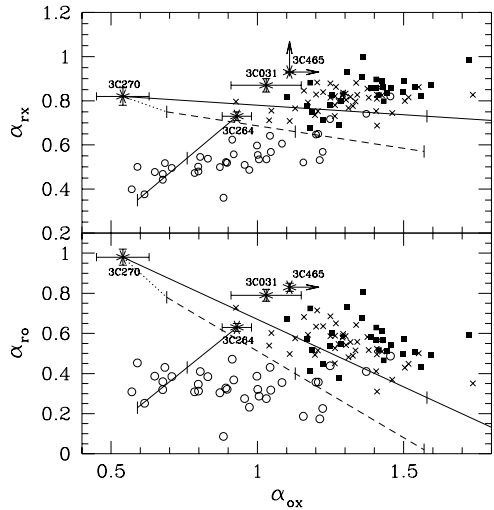,height=7.0truecm,angle=0}}
\caption{Broad band spectral indices of the FR~I nuclei (large stars)
compared to those of blazars (circles are XBL, crosses
and squares are RBL and FSRQ, respectively), as from Fossati et al. (1998). 
Continuous lines
represent the variations of spectral index due to the shift in energy
of the Doppler boosting. Ticks on the lines correspond to a increase of
$\delta$ by 10 and 100. The dashed line follows the beaming track of 3C~270
after a correction for absorption corresponding to $A_R = 2$.}
\end{figure}

\end{document}